# Model for self-consistent analysis of arbitrary MQW structures

Ivan Safonov[a*], Aleksey Shulika[a], Igor Sukhoivanov[a,b], Volodimir Lysak[a,c]
[a]Lab "Photonics" Kharkov National University of Radio Electronics,
Lenin av., 14, Kharkov, 61166, Ukraine;
[b]Departamento de Electronica, FIMEE, University Guanajuato,
Postal 215-A, Salamanca, GTO, 36730 Mexico;
[c]Gwangju Institute of science and technology,
1 Oryong-dong, Buk-ku, Gwangju, 500-712, Republic of Korea

## ABSTRACT

Self-consistent computations of the potential profile in complex semiconductor heterostructures can be successfully applied for comprehensive simulation of the gain and the absorption spectra, for the analysis of the capture, escape, tunneling, recombination, and relaxation phenomena and as a consequence it can be used for studying dynamical behavior of semiconductor lasers and amplifiers. However, many authors use non-entirely correct ways for the application of the method. In this paper the versatile model is proposed for the investigation, optimization, and the control of parameters of the semiconductor lasers and optical amplifiers which may be employed for the creation of new generations of the high-density photonic systems for the information processing and data transfer, follower and security arrangements. The model is based on the coupled Schrödinger, Poisson's and drift-diffusion equations which allow to determine energy quantization levels and wave functions of charge carriers, take into account built-in fields, and to investigate doped MQW structures and those under external electric fields influence. In the paper the methodology of computer realization based on our model is described. Boundary conditions for each equation and consideration of the convergence for the method are included. Frequently encountered in practice approaches and errors of self-consistent computations are described. Domains of applicability of the main approaches are estimated. Application examples of the method are given. Some of regularities of the results which were discovered by using self-consistent method are discussed. Design recommendations for structure optimization in respect to managing some parameters of AMQW structures are given.

## INTRODUCTION

Multiple quantum well heterostructures (MQWs) are the basis of modern optoelectronic devices. They are main elements of advanced planar waveguides, lasers, semiconductor optical amplifiers and photodetectors. The usage of MQWs allows to create compact integrated receiving and transmitting modules. The application of asymmetric twin-waveguide technology demonstrated in [1] allows designing the compact integrated receivers which contain SOA and p-i-n photodetector inside the single optical chip and which are capable of receiving data flow at bit rate 36 Gbit/sec.

The management of the properties of these devices is based on applying the multi-component solid solutions, the advanced methodologies of crystals growing and the usage of novel types of low-dimensional structures: quantum dot (QD), strained multiple quantum well (SMQW) and asymmetrical multiple quantum well (AMQW) heterostructures. In AMQW structures, each active region consists of multiple quantum wells of varying thickness and/or composition. If they are designed properly, AMQW structures have the potential to provide a spectral gain range more than twice as broad as in SMQW structures [2]. It was shown both theoretically and experimentally that QDs, SMQWs and AMQWs have a great potential to improve properties of SOAs [2-5]. Powerful device applications of such structures come from high coupling of quantum, optical, electrical, and transport phenomena inside them. Therefore theoretical investigation of such structures and devices based on them requires comprehensive physical models.

[*] e-mail safonov@kture.kharkov.ua; phone +380570-7021384; fax +380570-7021017

The coupling of variety physical processes in lowdimensional heterostructures which have to be taken into account, requires involving special schemes for the analysis of model equations using the so-called self-consistent approach. It was applied to the QW structures for the first time under investigation of the Hall effect in a thin film [6]. At present this approach is widely used for research of QW structures because it represents in the closest fit the interference of various physical processes.

In this paper a versatile model is proposed, which uses self-consistent approach. First we formulate the model. It is based on coupled the Schrödinger, Poisson's and drift-diffusion equations which allow to determine bound and quasi-bound states, take into account built-in fields, and investigate doped MQW structures and those under external electric fields. After that we review approaches and errors of self-consistent computations frequently encountered in practice. Then the boundary conditions for each equation are presented. As an example we show computation results for AMQW structure which uncover some interesting properties of the structure. We give design recommendations for the structure optimization based on these results in respect to managing some parameters of AMQW structures. Also we will discuss some results' regularities which were discovered by using the self-consistent method.

## 1. MODEL FOR SELF-CONSISTENT INVESTIGATION OF AMQW

### 1.1 Main equations of the model

Usually it is necessary to know the potential profile of the structure, position of quantum levels in a quantum-size region of the structure, the energy and the spatial carrier distribution to analyze main parameters and characteristics such as: voltage-current characteristic, gain spectrum, luminescence spectrum, capacity of the structure, etc. With these purposes we start with effective-mass 1D Schrödinger equation using the envelope function approximation for heterostructures. It is expressed in the form:

$$\left(-\frac{\hbar^2}{2}\frac{d}{dz}m_{c,v}(z)\frac{d}{dz} + V_{SC}(z)\right)\psi_{c,v,i}(z) = E_{c,v,i}\psi_{c,v,i}(z), \qquad (1)$$

where $\psi_{c,v,i}(z)$ is the envelope wave function for *i*-th quantization levels in the conduction and valence bands; $E_{c,v,i}$ is the energy of *i*-th quantization levels in the conduction and valence bands, $V_{SC}(z)$ is the self-consistent value of the potential energy, now it is unknown; $m_{c,v}$ is the effective masses of the charge carriers. Heavy-hole effective mass is defined as [7]:

$$m_v = \frac{m_0}{\gamma_1 - 2\gamma_2}, \qquad (2)$$

where $m_0$ is the electron mass; $\gamma_1$, $\gamma_2$ are the Luttinger's parameters which are determined from (10).

In the presence of the strain in the structure the equation (1) must be modified. Strain effects are taken into account through the modification of the Hamiltonian. The kinetic energy operator is supplemented with the dependence on the lattice constant and the potential energy operator is added by the term which gives a proper account of the deformation [8, 9]. The strain in QW layers is many-sided and it results in the interrelated changes in the structure. The shape and edges of the bands, effective masses and lattice constants are changed. Therefore, it is inadmissible to discard effects which are caused by these changes if it is necessary to achieve quantitative results.

The self-consistent potential energy operator is usually represented as a following sum:

$$V_{SC}(z) = V_{c,v}(z) + V_H(z) + V_{xc}(z), \qquad (3)$$

where $V_{c,v}(z)$ are the edges of conduction and valence bands respectively in bulk host-materials; $V_H(z)$ is the Hartree term of the potential energy due to non-uniform carriers distribution; $V_{xc}(z)$ is the exchange-correlation potential.

The exchange-correlation potential $V_{xc}(z)$ can be calculated in the framework of the local density approximation (LDA) in the Kohn-Sham density functional theory (DFT) [10, 11] or the generalized gradient approximation (GGA) [12].

To obtain the Hartree potential it is necessary to solve Poisson's equation:

$$\varepsilon_0 \frac{\partial}{\partial z} \varepsilon(z) \frac{\partial}{\partial z} \phi(z) = \rho(z), \qquad (4)$$

where $\varepsilon_0$ is the permittivity of the vacuum; $\varepsilon(z)$ is the relative permittivity of the material according to equation (9) (chapter 1.3). Because it is a piecewise-constant function it can be removed out the derivative. $\phi(z)$ is the Hartree potential; $\rho(z)$ is the charge density. In the presence of doping it can be calculated as the sum:

$$\rho(z) = -e_0 \left( p(z) - n(z) + N_d^i(z) - N_a^i(z) \right) \qquad (5)$$

where $p(z)$, $n(z)$ are the hole and electron concentrations; $N_d^i(z)$, $N_a^i(z)$ are the ionized donor and acceptor dopant concentrations respectively.

The ionized dopant concentration dependence on the temperature can be found by the following formula [8, 13]:

$$N_d^i(z) = \frac{N_d(z)}{1 + 2 \cdot e^{\frac{F_c + E_d}{k_B T}}}, \qquad N_a^i(z) = \frac{N_a(z)}{1 + 4 \cdot e^{\frac{F_v - E_a}{k_B T}}} \qquad (6a,b)$$

where $N_d(z)$, $N_a(z)$ are the donor and acceptor concentrations; $F_c$, $F_v$ are the Fermi quasi-levels for electrons and holes; $E_d$, $E_a$ are the ionization energies of donors and acceptors respectively; $T$ is the lattice temperature.

The spatial distribution of free charge carriers can be calculated if envelope wave functions of each quantization level and the low of carriers' energy distribution are known. Envelope wave functions can be obtained by solving the Schrödinger equation (1). Then spatial distribution of charge carriers is:

$$n(z) = \sum_{i=1}^{N} n_i |\psi_{c,i}(z)|^2, \qquad p(z) = \sum_{j=1}^{M} p_j |\psi_{v,j}(z)|^2 \qquad (7a,b)$$

where $n_i$, $p_j$ are the average values of electron and hole concentrations on $i$-th and $j$-th quantization levels respectively. They can be calculated by using the Fermi distribution. $N$ and $M$ are the number of quantization levels for electrons and holes. To include the quasi-bound states into consideration it is necessary to change the sums in equations (6a), (6b) on integrals because of continuity of the spectrum of quasi-bound states.

The coupled equations (1-7) allow to determine the eigen-states spectrum and the carrier distribution in a quantum-size region. But this system does not allow to investigate the influence of the external electric field. In the dielectric-like approach the external field is included in the boundary conditions of Poisson's equation and in the potential energy as the operator term which is linearly dependent on coordinate term [14]. In this case the semiconductor conductivity and generation-recombination processes are excluded.

Real devices usually have bulk layers and the influence of the charge carriers in them must be included into consideration. Determination of the carrier distribution in the bulk region is based on the solving of drift-diffusion equations coupled with continuity equations. The detailed description of the drift-diffusion calculation can be found elsewhere [15]. In general the equation system is as follows for the bulk region:

$$\frac{\partial n}{\partial z} = G - R - \frac{1}{e_0} \frac{\partial J_n}{\partial z}, \qquad \frac{\partial p}{\partial z} = G - R + \frac{1}{e_0} \frac{\partial J_p}{\partial z} \qquad (8a,b)$$

$$J_n = e_0 \mu_n n \frac{d}{dz}(V_H + V_n) + k_B T \mu_n \frac{dn}{dz}, \qquad J_p = e_0 \mu_p p \frac{d}{dz}(V_H - V_p) - k_B T \mu_p \frac{dp}{dz} \qquad (9a,b)$$

where $G$, $R$ are the rates of generation and recombination processes; $J_n$, $J_p$ are the electron and hole current densities; $V_n$, $V_p$ are the band parameters which includes band discontinuities [15, 16], $\mu_n$, $\mu_p$ are the carriers' mobilities.

Under the external electrical field the quantum-size region can not be described in the framework of the drift-diffusion model described above. For the determination of the non-equilibrium carrier distribution even in the case of the DC in the strict sense it is necessary to take into account transport processes in the context of the quantum theory. Here capture/escape, tunneling, relaxation, generation and different types of recombination processes must be taken into consideration. It is evident that including all the just-listed processes in the model implies the complication of the task. Therefore classic drift-diffusion equations are often used but they are modified for including the influence of some quantum processes [17].

**1.2 Assumptions and approximations**

Some approximations often used for self-consistent methodology of QWs investigation have been described in (chapter 1.1). Here we consider assumptions and approximations not mentioned above. A part of them is used in our model (approximations 1,3,4,10).

1) Effective mass approximation is widely used for solid-state devices modeling. It gives correct and matching to experiment results in most cases [18]. However, this approximation is obtained for completely periodical structure for known carriers' energy value. That is why the using of this approximation for thin-layer structures in a wide energy range can lead to inaccurate results [19]. The most part of the carrier spectral density $\partial n/\partial E$ falls usually at the energy range $(2 \div 3)k_B T$, i.e. not more than a few tens of meV. In this case we can suppose with a high accuracy that effective mass does not depend on carriers' energy. However, the large width condition is often unsatisfied. In a number of papers 2-4 atom layers structures are studied [20]. It is unacceptable without more detailed investigation.

2) Some types of charge carriers are excluded from the consideration. Many processes are really defined by only one type of carriers. For example the rate of the capture process for heavy holes is greatly higher than for electrons due to larger effective mass of heavy holes [21] so the total capture rate mainly depends on the electron capture process. But heavy holes are playing a main role while considering the influence of non-uniform carrier distribution on the potential profile in QW-region. This can be explained by better localization of high-mass particles in QWs. In a set of tasks when the mechanical stress is absent it is possible to neglect the light-holes and spin-orbit splitting [22] but full exclusion of carriers of any sign from the model is incorrect.

3) The heterojunction is abrupt. In the light of success of the QW structures growing technology and technique this approach is necessary to consider as an actual state of affairs. Nevertheless we should keep in mind the existence of image forces which make the heterojunction smoother. In comparison with other physical effects heterojunction smoothing is negligible. Therefore in practice this approach is close to the truth.

4) Dopants and components of the semiconductor alloy are uniformly distributed. Nowadays the level of the technology is enough to make very homogeneous doping [23]. So this approximation does not give us a large error. But earlier it was a cause of disagreement with an experiment [24, 25].

5) The Shottky approximation is often used under solving of the Poisson's equation on the metallurgical interface between a QW region and a bulk layer. This approximation implies complete absence of free carriers near the interface that is facilitated greatly solving and gives good result for metal-semiconductor contact and for contacts of heavy-doped semiconductors. The use of this approximation to consider contacts of intrinsic semiconductors or low-doped ones gives incorrect results.

6) The dopants are fully ionized. Sometimes this approximation is used in spite of the fact that the ionization level can be easily calculated (eq. 5). This approach does not introduce a large error at room temperature. However, it should be specified under not enough high temperatures.

7) Only the lowest quantization levels are under consideration. In a number of papers [8, 26] one considered a fixed number of levels because the population of the lowest levels is significantly higher. Moreover, it is possible to obtain an analytical relationship between Fermi level and carriers concentration under assumption that only one level is populated. If some quantization levels are very closely situated (it is possible in AMQWs or in structures consisting wide QWs), the excluding of the highest levels can lead to sizeable error at tunneling investigation and sometimes at self-consistent potential calculation. Influence of the highest levels on the tunneling processes can be described by the fact that the interaction of the highest levels' wave-functions is large. Especially strong influence of the highest levels can be expected for structures with resonant-tunneling pumping through the quasi-bound states.

8) Influence of quasi-bound states is neglected. Quantum-size layers are the centers of the free carriers capture and they can operate as a Bragg reflector for the material waves. So, the presence of quantum-size layers leads to the perturbation of charge carriers' distribution both in bound and quasi-bound states [27, 28]. The perturbation influences on the capture/escape processes.
9) QWs are non-interacting. This approach is right for structures with heavy barriers only. For structures with thin barriers this approach as will be depicted below leads not only to tunneling processes excluding from the model but to incorrect determination of the quantization levels position.
10) Exchange-correlation interaction is absent. Frequently authors do not mention the exchange-correlation interaction at the model describing, in another case authors refer on earlier works [frequently used ref. 29], assess segment of the exchange-correlation potential and exclude it from the model referring to small quantity. Notice, that [29] is dedicated to n-i-p-i structures investigation for which the exchange-correlation potential is negligible in comparison with the Coulomb potential. However, transferring this statement to low-doped or intrinsic MQW-structures can be a parent of an error, particularly at resonance tunneling investigation because even slight potential profile perturbation and *pro tanto* quantization levels deviation can give rise to a large tunneling conditions' changing [20, 30 and others].
11) Coulomb interaction of free charges is neglected. This approximation leads to one-electron problem and to refusing the core of the self-consistent approach. It can greatly simplify the model and often provides results with enough accuracy for qualitative and sometimes for quantitative investigation. But it is necessary to ground this approach for every specific task.

**1.3 Material parameters of compounds**

According to Vegard's law [31] some parameters of solid-state solutions can be determined as a linear interpolation of parameters of solution components. A nonlinear complimentary bowing item differs for different parameters and different components sets. For example it is equal to zero for the lattice constant but it can be considerable at the band gap energy calculation, especially for solutions of direct and indirect binary semiconductors [32]. During a long time there were not unfounded disputes because this law was in contradiction with experiments [24, 25]. Later it was ascertained that this disagreement was conditioned by the non-ideality of the grown structures, especially by the non-uniform components distribution. With the technology development the parameters deviations from the Vegard's law for the zinc blend-like structures became less. This fact allows to make even quantitative estimations [33]. According to [34] the error of the Vegard's law applying is not higher than a percent. Interpolation equation for the multiple component solid solutions with any given number of components of $A^{III}B^V$ type can be written as follows:

$$a_{A_{x1}...A_{xk}B_{y1}...B_{yl}} = \frac{1}{\sum_i x_i \cdot \sum_j y_j} \sum_i \sum_j x_i y_j a_{A_i B_j}, \qquad (10)$$

where $a_{A_{x1}...A_{xk}B_{y1}...B_{yl}}$ is the under calculation parameter for the solution with the gross-formula $A_{x1}...A_{xk}B_{y1}...B_{ym}$ (it can be the effective mass of electrons, band gap energy, electron affinity, etc.), where $A_{x1}...A_{xk}$ are the elements of the III group, $B_{x1}...B_{xk}$ – are the elements of the V group; $a_{A_i B_j}$ are the binary $A_i B_j$ semiconductors parameters.

There are more precise higher order formulas. However, they are attached to the fixed multiple component solution and are not so universal [for example 35]. Besides, some parameters can be obtained from complex high-accurate methods based on the crystalline structure of the material.

**1.4 Iterative procedure**

The process of the equations solving is started with Schrödinger equation (1). Here it is assumed that charge carriers are uniformly distributed and terms of the potential energy operator in (2), which is caused by the Coulomb $V_H(z)$ and exchange-correlation $V_{xc}(z)$ interactions, are equal to zero. As a result of the solving the eigen-states spectrum is obtained. Then taking into consideration the Fermi distribution the carriers spatial and energy distribution is defined by using eqs. (6a,b). With obtained carriers distribution within the quantum-size region the potential profile and the position of quasi-Fermi levels is improved by the coupled solving of drift-diffusion (7a,b), continuity (8a,b) and Poisson's (3) equations. Then the Schrödinger equation is resolved using improved quantities of $V_{SC}(z)$, $F_c$ and $F_v$. This iterative solving is carried out till the potential profile changing from iteration to iteration is larger than the required quantity. The number of iterations depends on the nonuniformity of carrier distribution and may be from 5–6 to 30 or more for given computational error of 1%. It does not allow to set the fixed number of iterations.

**1.5 Boundary conditions**

Correct approximations are not enough for accurate solution of a problem. We need to set boundary conditions for each equation.

For Schrödinger's equation (2) the zero boundary conditions are set, i.e. wave functions and their first-order derivatives should be zero on the edges of the structure. These conditions are usually approximated by enough thick wide-gap cladding layers. It is correct if we consider bound states only. For the quasi-bound it is not correct because for these states one or both sides of the structure are not fully reflected (a state is not fully bound). In this case boundary conditions which take into account the existence of quasi-bound states should be used [36, 37]. For the reflected and not reflected edges boundary conditions may be written as plane waves [36]:

$$\psi_{c,v,i}^{r}(z) = I_{c,v} e^{i\,\text{sgn}(m_{c,v})k_{c,v,i}z} + r_{c,v} e^{-i\,\text{sgn}(m_{c,v})k_{c,v,i}z}, \qquad \psi_{c,v,i}^{non-r}(z) = t_{c,v} e^{i\,\text{sgn}(m_{c,v})k_{c,v}z}, \quad (11a,b)$$

In our opinion boundary conditions for drift-diffusion equations and continuity equation may be set correctly by requirement of charge neutrality of bulk claddings and continuity of currents and their first-order derivative on the edge of the structure [38, 39].

## 2 PROTOTYPE STRUCTURE

As it has been mentioned above AMQWs have a great potential to improve properties of SOAs and may be used for amplification of ultrashort optical pulses by duration of tens femtoseconds [40, 41]. The prototype structure is composed with 3 pairs of narrow gap layers which are separated by wide-gap ones. The width of all layers is 5 nanometers. Our computations have shown [41] this structure has potential to obtain the gain bandwidth more than 100 nanometers. The composition of the structure is shown in the table.

Table 1

| Layer(s) | Cladding | Barriers, separate confinement | 1st, 2nd QWs | 3rd, 4th QWs | 5th, 6th QWs |
|---|---|---|---|---|---|
| Material | InP | $In_{0.68}Ga_{0.32}As_{0.7}P_{0.3}$ | $In_{0.82}Ga_{0.13}As_{0.99}P_{0.01}$ | $In_{0.8}Ga_{0.2}As_{0.95}P_{0.05}$ | $In_{0.72}Ga_{0.28}As_{0.95}P_{0.05}$ |

**2.1 Geometry influence**

The influence on the eigen-states spectra of the widths of the barriers and QWs is clear in the case of flat bands and one kind of charge carriers (ambipolar approximation). The convenience of these approximations lies in their simplicity. They can reflect an actual state of affairs precisely enough in single cases. They are threshold states of QW laser structures with rather narrow optical confinement layers (OCL). However, flat bands and ambipolar approximation are unacceptable in all other cases because of regions of charge neutrality violation, the sizes of which result in bending of the band edges. Here we will consider a purely dimensional AMQW structure keeping the elements of symmetry, which are peculiar to the prototype structure. All of QWs will be identical in width. Barriers also will have the same widths which are different from QWs. Computation results for structures with different width of barrier layers are shown in Fig. 1. Results for structures composed with narrow-gap layers with different width are shown in Fig. 2. Figures 1 and 2 show electron eigen-state spectra in conduction band with taking electron-hole Coulomb interaction into account. Similar results may be obtained for other carriers.

As it is shown in Fig. 1 the variation of width of barriers is a powerful instrument for control of quantizing levels position. The following effects which are concerned with the barrier width variation may be tracked out.

The distance between levels of the states of coupled identical QWs changes appreciably. The change for electrons can be so sizeable that levels of different pairs of QWs could change their order. This is observed in the case of 2 nm width barriers. Indeed, if the difference in the depth of QWs of first and second groups consists of $5.18 \cdot 10^{-21}$ J (32.3 meV) the change of spacing between levels of the coupled states of first-group QWs will be $8.08 \cdot 10^{-21}$ J (50.4 meV) for lowest levels under switch from 5-nm-barriers to 2-nm-barriers. This is enough for the violation of the initial sequence order of quantized levels even without taking into account the change of level spacing of second-group QWs. Such a change of the lowest level position results in the change of the minimal frequency of the optical transition more than 2.45 GHz (without taking into account displacement of hole levels as it is much smaller). In the wavelength range close to 1.55 μm such a frequency change means wavelength diminution approximately on 20 nm. If we take into account the fact that the channel spacing in modern DWDM systems can be less than 1 nm [42], then it becomes clear that such a wavelength variation is not negligible.

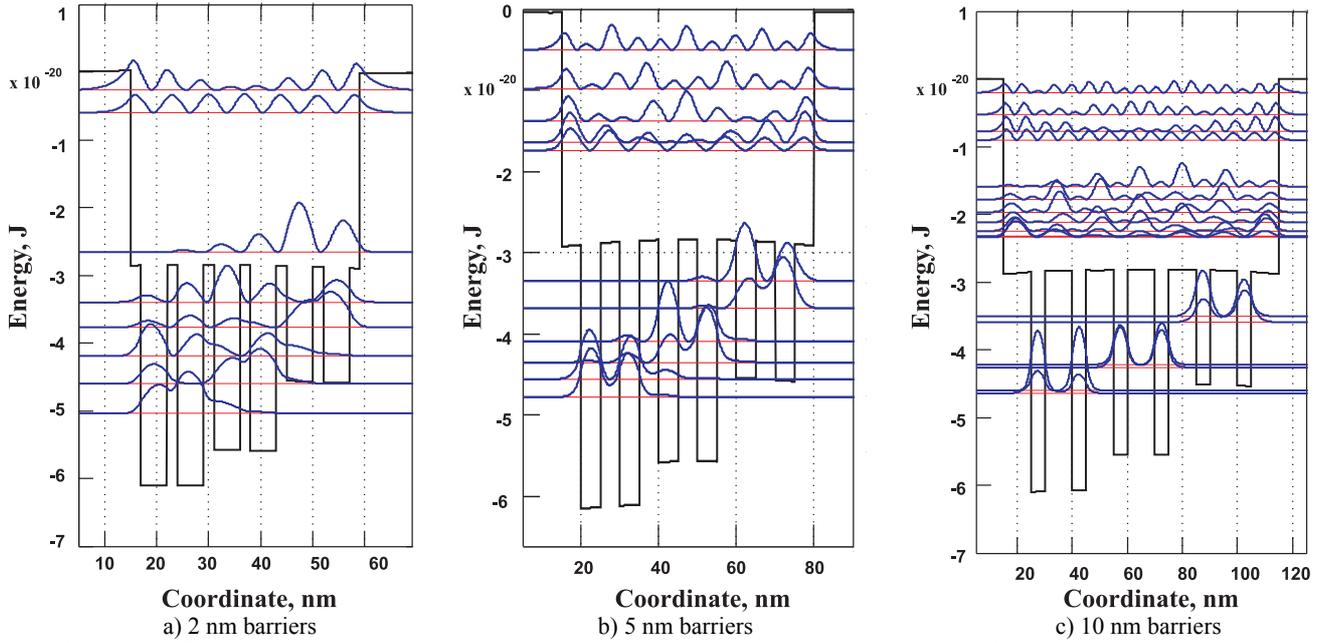

Figure 1 – Influence of the barriers width on the electron eigen-states in AMQW. The QW width is equal to 5 nm for all cases

We note that the change of the levels position under variation of the barriers width is determined mainly by the width variation of those barriers which separate identical QWs. The change of the barriers width between groups of wells governs only a degree of coupling for different-group QWs.

The second effect is concerned with the change of the wave functions localization. For the case of thick barriers the tails of wave functions which correspond to states of QWs from one group almost do not penetrate to other QWs. The penetration is significant in the case of thin barriers. On the one hand this effect results in more uniform carrier distribution in the structure. On the other hand the strong coupling leads to significant transformation of the spectral density of states and as a consequence to transformation of gain and absorption spectra [43]. Estimations show [43] that great distinction from the case of uncoupled QWs will be observed if $\Delta E \geq 2\hbar/\tau_{in}$, where $\Delta E$ is the spacing between adjacent levels, $\tau_{in}$ is the intraband relaxation time. The holes are strongly localized even for 2 nm barriers because of larger effective mass.

The next effect is concerned with the fact that the increasing of barrier widths leads to better separation of regions with the positive total charge and the negative total charge. The separation is reflected in the appearance of local maxima and minima of Hartree potential. But in this case they are not visible in the scale of the figure.

Another effect which we can see is the change of a number of states in the separate confinement region above QWs. The main factor is a variation of the structure width as a whole.

Note the Hartree potential difference changes along the structures with different barriers is not large because it is generally defined by the non-equal penetration of electrons and holes into the cladding layers.

Additional features can be obtained by the QWs' width varying. In Fig. 2 the first effect is the increasing of the quantization levels number with the QW widening. But in this case in contrast to barriers widening the levels number increased not only in the confinement region but within wells too. With the QW width decreasing the minimal energy of optical transition is increased it is so-called blue shift of the gain spectrum edge. Minimal energies of optical transitions are 1831 nm, 1676 nm and 1536 nm for 3 nm (Fig. 2a), 5 nm (Fig. 2b) and 10 nm (Fig. 2c) QWs correspondingly (without taking into account levels widening). The nature of the levels shift is appreciably different with a case of barriers width varying. The shift for barriers is determined by the changing of QWs interaction conditions. In this case the levels of coupled identical wells position changes not around the certain energy center which is the position of solitary well level, but this center changes the position. In other words in a case of barriers varying part of levels is shifted up and a part is shifted down from the level splitting. And at QW widening all levels are shifted down.

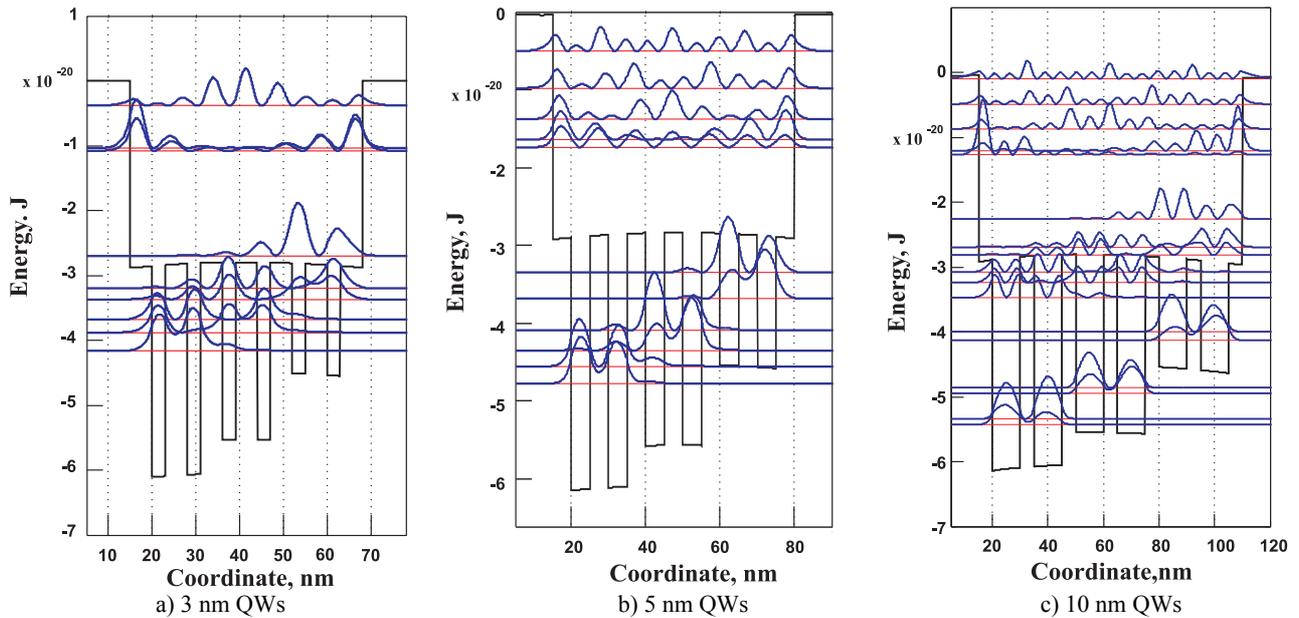

a) 3 nm QWs     b) 5 nm QWs     c) 10 nm QWs

Figure 2 – Influence of the QWs width on the electron eigen-states in AMQW. The barrier width is equal to 5 nm for all cases

The next observed effect is the potential peaks of the Hartree potential appearance for each well with QWs widening. But this effect development is lower than at barriers widening. It is related with better spatial separation of charges with different signs.

We conclude that at a small total width of AMQW structure the perturbation of the potential profile which is induced by the non-uniform distribution of the charge carriers is negligibly small. In this case the carrier induced field significantly influences upon the tunneling of the heavy holes because of much greater mass. If we are not interested in heavy hole tunneling the potential profile of the narrow AMQW structure can be approximated by the piece-wise constant (or piece-wise linear for the biased structure) band diagram and the self-consistent approach is superfluous.

**2.2 Symmetry influence**

Structures shown in chapter 2.1 are asymmetric in respect to the solid-state composition. But they have geometric symmetry. The symmetry of such structures lies in the equality of width of QWs and/or barriers and in the equal by pairs of QWs depth. Any varying of QWs depth does not uncover new structure features, excluding some special cases[*]. At the same time varying of the layers width allows to investigate new effects.

Some examples of geometrical asymmetries are shown in the Fig. 3. The data about width of barrier and QW layers are tabulated below. In the structure (a) compositionally identical and consequently depth identical QWs have different width. In the structure (b) barriers which separate identical QWs are reduced. In the structure (c) each barrier layers have different width.

In the structure (d) in contrast to the structure (b) barrier width between identical QWs is reduced but between different QWs it is differently enlarged. Finally, the feature of the structure (e) is a large width of separate confinement layers. In the Fig 3a it is shown that the compositional coincidence of the QWs does not lead to the relation between them if their width is different. Adjacent QWs almost do not influence on the position of quantization levels. It is possible to get cases for which interaction of farther QWs is higher than of adjacent QWs. For example square of the wave-function absolute value for fifth from below quantization level is in the third QW. At the same time the maximum probability of the electron detection in other well is not within the second or fourth well but within the sixth well. In the case of structure (a) only one level corresponds to each QW but for wider well a number of levels can be larger. In this situation it is possible to obtain the interaction of different QWs by only one level. Apparently it is possible to build structures with more than two QWs which differ in the width and depth but coupled by one quantization level. In such structures tunnel-resonance current pumping is suitable.

---

[*] They are cases when the wave-functions which have different number of zeros are strongly interacting. It allows to synthesize structures for the second harmonics generation [44].

Table 2

| Structure | Layer | Left separate confinement | 1st well | 1st barrier | 2nd well | 2nd barrier | 3rd well | 3rd barrier | 4th well | 4th barrier | 5th well | 5th barrier | 6th well | Right separate confinement |
|---|---|---|---|---|---|---|---|---|---|---|---|---|---|---|
| a) | Width, nm | 5 | 3 | 5 | 5 | 5 | 3 | 5 | 5 | 5 | 3 | 5 | 5 | 5 |
| b) | | 5 | 5 | 5 | 5 | 2 | 5 | 5 | 5 | 2 | 5 | 5 | 5 | 5 |
| c) | | 5 | 5 | 4 | 5 | 2 | 5 | 3 | 5 | 7 | 5 | 2 | 5 | 5 |
| d) | | 10 | 5 | 3 | 5 | 10 | 5 | 3 | 5 | 15 | 5 | 3 | 5 | 10 |
| e) | | 25 | 5 | 3 | 5 | 3 | 5 | 5 | 5 | 3 | 5 | 5 | 5 | 25 |

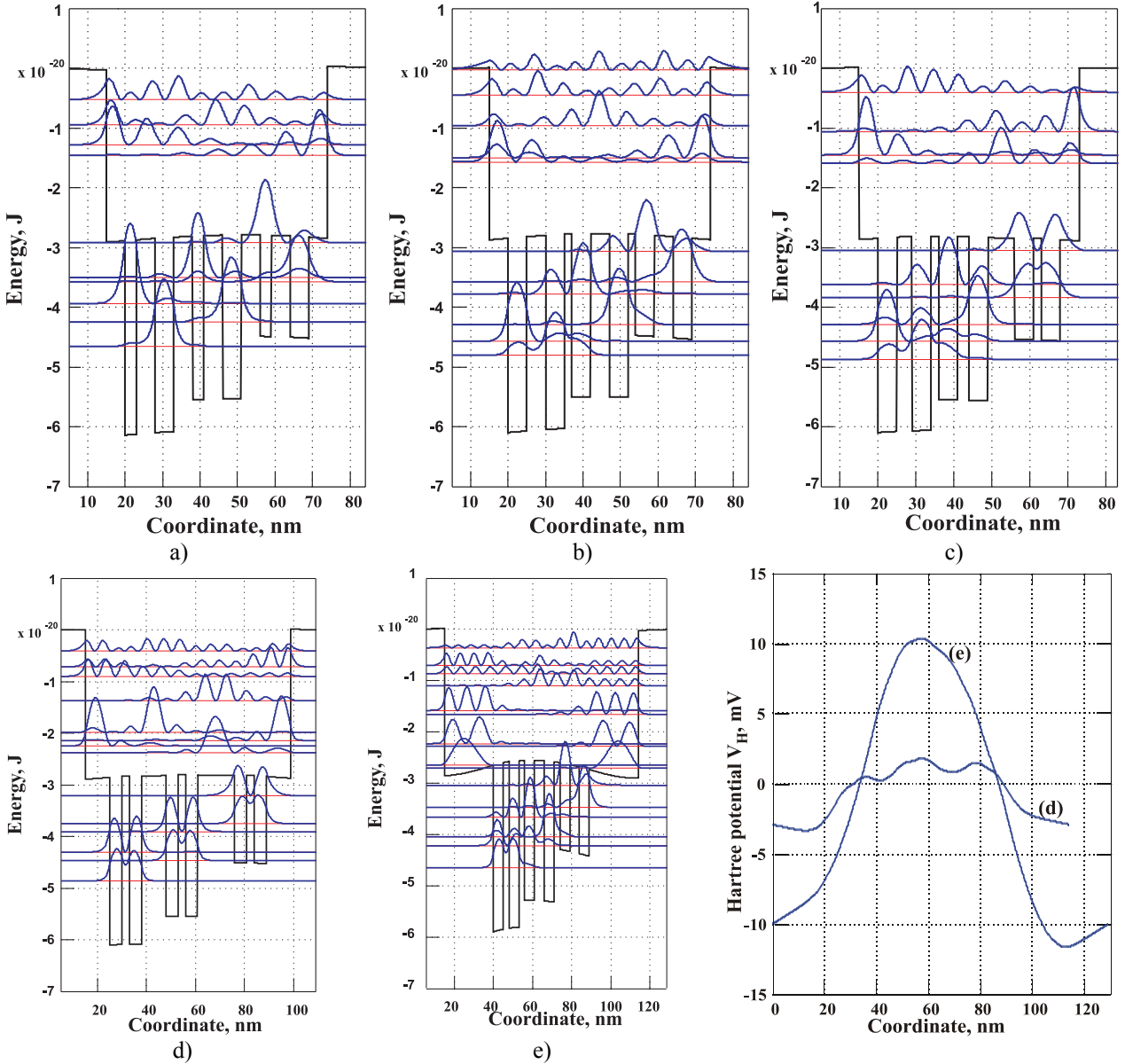

Figure 3 – Symmetry influence on the electron eigen-states in AMQW and on the potential profile

Figure 4 – Hartree potential for structures (d) and (e)

Fig. 3b uncovers the fact that by the decreasing of the barriers between identical QWs it is possible to get strong interaction of wells of different groups. This fact also can be used for tunnel pumping. An interesting result can be obtained from the comparison of quantization levels position in the prototype structure and in the structure in the Fig. 3b. The position of two lowest levels almost has not been changed. The third level is moved more and displacement of the highest levels is very large (the value of the displacement is comparable with the energy distance between levels). The explanation of this fact requires the additional investigation. However, we suppose that two factors can be a root of such a behavior. There is the increase of the interaction of QWs of different groups and the decrease of the total width of QW-region. Probably the first factor is principal. It is because the lowering of levels occurs when deeper QWs interact with less deeper one. Such an explanation allows to understand a small displacement of the lowest levels. This is because the deeper QWs interact with less deeper one not so strong. For wide wells (see Fig.3c) the lowest levels can be situated in the band gap of less deep wells.

Figs. 3c,d demonstrate the fact of charge carriers localization in the confinement region and this localization is stronger for the lowest barrier states. In spite of this maximums of the Hartree potential are within QWs (Fig. 4). This is because of larger probability population of the lowest QW states. The weak separation of charges inside the each group of QWs is the cause of the fact that potential maximums appear not inside the each single QW but inside the each group of QWs.

The result which is strongly differs from the previous ones is shown in Fig. 3e. The influence of the Hartree potential in this structure is much higher. The build-in potential difference induced by the non-uniform carrier distribution is about 22 mV (see Fig. 4). This value is as much as 3 times higher of such a potential for the prototype structure. At the same time the separation of charges of the different signs between the barriers and QWs is greatly smaller that charge separation between OCL and QW-region. It is manifested by absence of the local maxima of $V_H$ in Fig. 4.

**2.3 Recommendation for AMQW design**

It was shown above the variation of the layers width allows to manage the eigen-states spectrum and as a consequence managing macroscopic parameters and characteristics of a device as a whole. Transport and spectral properties of AMQW-devices are defined by the eigen-states spectrum and spatial and energy carriers distribution.

The order of the geometry fitting we propose is the following. First in rough approximation the number and the composition of layers are selected. Then the structure with the closest fit of quantization levels position to the necessary one is produced by the variation of widths of the narrow-band layers. The precision of the varying is not higher than the atomic layer. This accuracy can be not enough to obtain the necessary eigen-states spectrum. More precise correction of levels position and managing the carrier space distribution is possible by variation of the barriers widths. Note, the width and the relative position of barriers have greater influence on the carrier wave-functions, energy of which exceeds the barrier height. The influence of the width and the relative position of QWs is less. As a consequence geometry of the barriers has greater influence on the carrier capture conditions. The case of wide and especially identical barriers leads to high localization of carriers in barrier layers (figs. 3d,e) and to the decreasing of the capture efficiency.

Thus, the frequency position of the gain spectrum is generally defined by the QWs' geometry but the shape of the spectrum, its dependence on the pumping current and rate parameters of a device should be generally defined by the barriers geometry. Under other identical conditions the structure with thinner barriers is preferable in respect to potential profile perturbation, capture and tunneling efficiency.

## CONCLUSION

Complex semiconductor low-dimensional heterostructures, such as asymmetrical and strained multiple quantum wells, quantum dots, dots-in-a-well structures have great potential to improve properties of lasers and amplifiers. The coupling of variety physical processes in these structures, which should be taken into account requires complex models and the involving of special schemes for analysis of equations of a model. Such a model is proposed in this paper and computation results for AMQW amplifier structure are presented.

The model allows to investigate multilayer structures with arbitrary configuration of layers with doping and without it and to take external electric field into account. The ways to the correct account for the influence of transport processes on the potential profile are indicated. Main approximations are described as a source of mistakes under simulation. Computational results for a wide gain band AMQW structure are presented based on the model. Basing on these results we have generated recommendations for synthesis of arbitrary semiconductor AMQW structure with necessary characteristics. Later we will use this model for more precise computations of the capture, escape and tunneling times in complex MQWs.

## ACKNOWLEGMENTS

Authors acknowledge Klimenko M. V., Mashoshyna O. V., and Bogdan O. N. (Lab. "Photonics", Kharkov National University of Radio Electronics, Kharkov, Ukraine) for fruitful discussions and useful advice during the work fulfillment and writing the paper.